\documentclass[]{spie}  %>>> use for US letter paper
\usepackage{astro_bib_macro} %raccourcis noms de journaux d?finis dans ce fichier

\usepackage{amsmath,amsfonts,amssymb}
\usepackage{mathtools}
\usepackage{graphicx}
\usepackage{bm}% to write bold letters in math mode	
\usepackage[colorlinks=true, allcolors=blue]{hyperref}
\title{Apodized Pupil Lyot Coronagraphs with arbitrary aperture telescopes: novel designs using hybrid focal plane masks}

\author{Mamadou N'Diaye\supit{a},
Kevin Fogarty\supit{b},
R\'{e}mi Soummer\supit{b}, 
Alexis Carlotti\supit{c}, 
Kjetil Dohlen\supit{d},\\ 
Johan Mazoyer\supit{e}, 
Laurent Pueyo\supit{b}, 
Kathryn St.Laurent\supit{b}
Neil Zimmerman\supit{f} 
\skiplinehalf
\supit{a} Universit\'{e} C\^{o}te d'Azur, Observatoire de la C\^{o}te d'Azur, CNRS, Laboratoire Lagrange, Bd de l'Observatoire, CS 34229, 06304 Nice Cedex 4, France\\
\supit{b} Space Telescope Science Institute, 3700 San Martin Drive, Baltimore, MD 21218, USA\\
\supit{c} Universit\'{e} Grenoble Alpes, IPAG, 38000 Grenoble, France\\
\supit{d} Aix Marseille Universit\'{e}, CNRS, LAM (Laboratoire d\'{}Astrophysique de Marseille) UMR 7326, F-13388, Marseille, France\\
\supit{e} Johns Hopkins University, 3400 North Charles St, Baltimore, MD 21218, USA\\
\supit{f} NASA Goddard Space Center, 8800 Greenbelt Rd, Greenbelt, MD 20771, USA
}

\authorinfo{Further author information, send correspondence to Lucie Leboulleux: E-mail: mamadou.ndiaye@oca.eu}

\begin{document} 
\maketitle

\begin{abstract}
Exoplanet imaging and spectroscopy are now routinely achieved by dedicated instruments on large ground-based observatories (e.g. Gemini/GPI, VLT/SPHERE, or Subaru/SCExAO). In addition to extreme adaptive optics (ExAO) and post-processing methods, these facilities make use of the most advanced coronagraphs to suppress light of an observed star and enable the observation of circumstellar environments. The Apodized Pupil Lyot Coronagraph (APLC) is one of the leading coronagraphic baseline in the current generation of instruments. This concept combines a pupil apodization, an opaque focal plane mask (FPM), and a Lyot stop. APLC can be optimized for a range of applications and designs exist for on-axis segmented aperture telescopes at $10^{10}$ contrast in broadband light. In this communication, we propose novel designs to push the limits of this concept further by modifying the nature of the FPM from its standard opaque mask to a smaller size occulting spot surrounded by circular phase shifting zones. We present the formalism of this new concept which solutions find two possible applications: 1) upgrades for the current generation of ExAO coronagraphs since these solutions remain compatible with the existing designs and will provide better inner working angle, contrast and throughput, and 2) coronagraphy at $10^{10}$ contrast for future flagship missions such as LUVOIR, with the goal to increase the throughput of the existing designs for the observation of Earth-like planets around nearby stars.
\end{abstract}


\section{INTRODUCTION}
\label{sec:intro}
High-contrast imaging and spectroscopy is a burgeoning field in astronomy with the observation of brown dwarfs, extrasolar planets, and circumstellar disks to advance in our understanding of the nature, formation, and evolution of exoplanetary systems. On-going progress in instrumentation has recently led to the deployment of high-contrast facilities\cite{Beuzit2008,Macintosh2014,Jovanovic2015} with extreme adaptive optics (ExAO) systems on the ground to directly image and spectrally analyze warm gas giant planets around nearby stars\cite{Macintosh2015,Chauvin2017}. Currently under construction, Extremely Large Telescopes (ELTs) will be coupled with instruments to enable the spectral characterization of the known exoplanets and possibly the detection of colder gaseous planets or massive telluric companions. In space, WFIRST\cite{Spergel2015} and several post-JWST mission concepts such as LUVOIR\cite{Bolcar2017} or HabEx\cite{Mennesson2016} are currently under study with the aim to observe planetary companions in reflected light down to exo-Earths.  

The emergence of WFIRST in the astrophysical landscape with its unfriendly pupil has favored the fast-growing development of high-contrast instrumentation over the past few years in all its aspects (wavefront sensing and control, coronagraphy, post-processing). In coronagraphy\cite{Guyon2006,Mawet2012}, many concepts have indeed come up to derive solutions that are suitable to any telescope aperture geometry, including secondary mirror central obstruction, spider struts, or primary mirror segmentation, while simultaneously providing: (i) high contrast to study the faintest planets in the vicinity of bright sources, (ii) high throughput to maximize the number of extracted photons from the observed planet, (iii) small inner working angle (IWA) for the observation of the planets at the closest separations from a given nearby star, and (iv) wide spectral band to enable the analysis of the planet atmosphere composition.

Among the numerous solutions in the literature, the Apodized Pupil Lyot Coronagraph (APLC)\cite{Aime2002, Soummer2005} is one of the leading concepts in the exoplanet imaging field with devices that are installed in the ground-based exoplanet imagers Palomar/P1640, Gemini Planet Imager (GPI), and VLT/SPHERE\cite{Hinkley2011,Beuzit2008,Macintosh2014}. The GPI version has parts that were designed to offer $10^7$ contrast at 5 resolution elements from a given observed nearby star over 20\% spectral band\cite{Soummer2009,Soummer2011}. This starlight suppression system operates routinely within these exoplanet imagers to survey hundreds of nearby stars and search for planetary companions.

Novel solutions have recently emerged for the observation of Earth twins with future large missions, enabling to produce a star image with $10^{10}$ contrast Point Spread Function (PSF) dark zone over 10\% bandpass with arbitrary telescope apertures, although with moderate IWA ($\sim$ 4 resolution elements)\cite{N'Diaye2015a,N'Diaye2016,Zimmerman2016}. Additionally for the past two years, the Segmented Coronagraph Design and Analysis (SCDA) study has been a coordinated on-going effort supported by NASA Exoplanet Exploration Program (ExEP) which has enabled to increase our understanding in the performance and limitations of coronagraph and overcome their current limits, see for example the case of the vortex coronagraph\cite{Ruane2017}. In the case of APLC, a large review of these works is given by St.Laurent et al., this conference\cite{st.laurent2018}

However, these APLC designs remain partially limited in terms of IWA, bandpass, and throughput, preventing the observation of close-in planets and the atmospheric characterization, and requiring long exposure time even with large aperture to collect enough photons from reflected light planets in the context of space applications. New solutions will allow us to enlarge planet discovery space and increase the spectral coverage of the APLC concept to study the nature of the close-in planetary companions. 

In this communication, we revisit the APLC concept by elaborating novel design optimization strategies and discussing different coronagraph mask geometries to reduce its IWA further.  In particular, we investigate the combination of the APLC with a related coronagraph concept, the dual zone phase mask (DZPM)\cite{Soummer2003b,N'Diaye2012a,Delorme2016} where broadband coronagraphic nulling is achieved by the destructive interference between non-$\pi$ phase delays. While the required performance does not appear to be achievable by this coronagraph alone, this paper discusses the use of the dual-zone concept in combination with the APLC. We finally present preliminary results to overcome the current limitations of the APLC and discuss possible applications for exoplanet studies with ground-based and space observatories.

\section{Principle of APLC/Shaped Pupil coronagraph with hybrid focal plane masks}
\label{sec:principle}

\subsection{Formalism}
The formalism of the Apodized Pupil Lyot Coronagraph (APLC) was described in several publications over the past few years. We briefly recall the main lines and refer the reader to the different publications in the references for a more detailed description \cite{Aime2002,Soummer2003a,Soummer2005,Soummer2009,Soummer2011,N'Diaye2015a,N'Diaye2016}. In the following, the vectors $\bm{r}$ and $\bm{\xi}$ represent the two-dimensional vectors in the pupil and focal planes, $D$ denotes the aperture diameter, $\lambda$ defines a given wavelength within the spectral bandwidth $\Delta\lambda$ centered around the wavelength $\lambda_0$. We denote $\hat{f}$ the Fourier transform of $f$. The $*$ symbol represents the convolution operator.  

\begin{figure}[!ht]
\centering
\includegraphics[width=\textwidth]{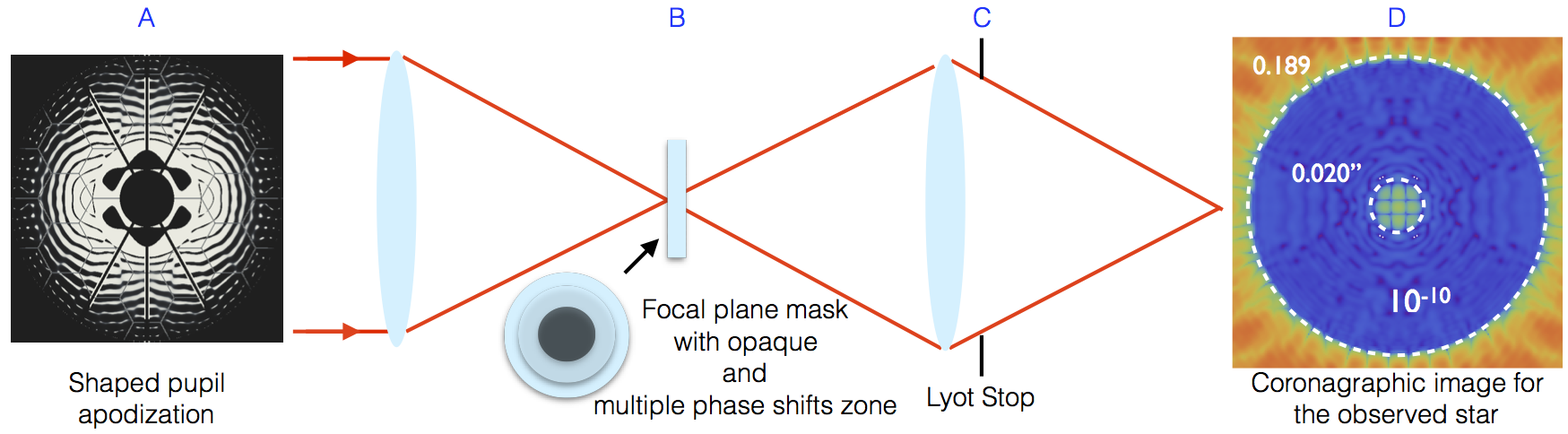}
\caption{APLC optical layout with an apodization in the entrance pupil plane A, a focal plane mask in the following focal plane B, a Lyot stop in the relayed pupil plane C, and the final image plane D in which the coronagraphic image is formed. In this schematic representation, the aperture is represented with a shaped pupil-type apodization for a segmented pupil. The focal plane mask is here made up with a central opaque part surrounded by two phase shifting annuli. The coronagraphic image on the right displays a blue annulus representing the $10^{10}$ contrast region, a.k.a. dark zone, in which the planetary companions are expected to be observed.}
\label{fig:APLC_layout}
\end{figure}

This concept involves four different plane in its optical layout: the entrance pupil plane A with the telescope aperture $P_0$, the focal plane B in which the focal plane mask (FPM) is located, the relayed pupil plane C in which the Lyot stop takes place to remove the diffraction effects due to the FPM, and the image plane D where the detector is placed. A schematic representation of the optical layout is given in Figure \ref{fig:APLC_layout}. 

The APLC combines an entrance pupil apodization $\Phi$, a focal plane mask of size $m$ and transmission $1 - M(\bm{\xi})$ with $M(\bm{\xi})=1$ for $\xi<m/2$ and 0 otherwise, and a Lyot stop $L$ to form the image of an observed nearby star with reduced intensity and enable the observation of planetary companions in orbit. We denote $ID$ and $OD$ the Lyot stop inner and outer diameters and express them in fraction of $D$. In our studies, we assume a coronagraphic setup such that there is no pupil magnification and the electric field between two successive planes are related by a Fourier transform.

The transmission of the pupil $P$ is then given by
\begin{equation}
    P(\bm{r})=\Phi(\bm{r})\,P_0(\bm{r})\,.
    \label{eq:pupil}
\end{equation}

Assuming an opaque hard-edge FPM, the electric field $\Psi$ is described in each plane and at the observing wavelength $\lambda$ by the following equations:
\begin{subequations}
      \renewcommand{\theequation}{\theparentequation\alph{equation}}
\begin{alignat}{2}
& \Psi_A(\bm{r}, \lambda)=P(\bm{r}), \label{subeq:Psi_A}\\
& \Psi_B(\bm{\xi}, \lambda)= \frac{\lambda_0}{\lambda}\widehat{\Psi}_A \left (\frac{\lambda_0}{\lambda}\bm{\xi}, \lambda \right ) \left (1- M(\bm{\xi})\right ), \label{subeq:Psi_B}\\
& \Psi_C(\bm{r}, \lambda)= \left (\Psi_A(\bm{r}, \lambda) - \Psi_A(\bm{r}, \lambda)* \frac{\lambda_0}{\lambda}\widehat{M}\left (\frac{\lambda_0}{\lambda}\bm{r}, \lambda \right ) \right )L(\bm{r}) \label{subeq:Psi_C},\\
& \Psi_D(\bm{\xi}, \lambda)= \left (\frac{\lambda_0}{\lambda}\right )^2 \left (\widehat{\Psi}_A\left (\frac{\lambda_0}{\lambda}\bm{\xi}, \lambda \right ) (1- M(\bm{\xi})) \right )*\widehat{L}\left (\frac{\lambda_0}{\lambda}\bm{\xi}, \lambda \right ). \label{subeq:Psi_D}
\end{alignat}
\end{subequations}

Although given for an opaque FPM, the formalism can be extended to a more complex geometry mask. In this communication, we investigate the use of an hybrid FPM, an opaque dot with size $m$ that is surrounded by two phase shifting annuli with sizes $m_1$ and $m_2$. Adding these two additional zones gives us additional degrees of freedom in the framework of the APLC to find solutions with reduced inner working angle (IWA $< 3 \lambda/D$) or large spectral bandwidth for the observation of exo-Earths with future space observatories. The amplitude transmission of such an hybrid mask writes as
\begin{equation}
    t = 1 - e^{i\varphi_1}M - (e^{i\varphi_2} - e^{i\varphi_1})\,M_1 - (1 - e^{i\varphi_2})\,M_2,
\end{equation}
in which $M$, $M_1$, and $M_2$ define the top-hat functions of the opaque dot, the inner and outer phase-shifting parts of the hybrid mask; they are equal to $1$ for $\xi < m/2$, $\xi < m_1/2$, and $\xi < m_2/2$, respectively, and $0$ otherwise. Assuming such a mask, the electric field $\Psi$ in each plane and at each wavelength writes as 
\begin{subequations}
      \renewcommand{\theequation}{\theparentequation\alph{equation}}
\begin{alignat}{2}
\Psi_A(\bm{r}, \lambda) &=P(\bm{r}), \label{subeq:Psi_A2}\\
\Psi_B(\bm{\xi}, \lambda) &= \frac{\lambda_0}{\lambda}\widehat{\Psi}_A \left (\frac{\lambda_0}{\lambda}\bm{\xi}, \lambda \right ) \left (1 - e^{i\varphi_1}M(\bm{\xi}) - (e^{i\varphi_2} - e^{i\varphi_1})\,M_1(\bm{\xi}) - (1 - e^{i\varphi_2})\,M_2(\bm{\xi})\right ), \label{subeq:Psi_B2}\\
\begin{split}
\Psi_C(\bm{r}, \lambda) &= \Psi_A(\bm{r}, \lambda)L(\bm{r})\\
&- \frac{\lambda_0}{\lambda}\Psi_A(\bm{r}, \lambda)*\left ( 
 e^{i\varphi_1}\widehat{M}\left (\frac{\lambda_0}{\lambda}\bm{r}, \lambda \right )
+ (e^{i\varphi_2} - e^{i\varphi_1})\widehat{M_1}\left (\frac{\lambda_0}{\lambda}\bm{r}, \lambda \right )
+ (1 - e^{i\varphi_2})\widehat{M_2}\left (\frac{\lambda_0}{\lambda}\bm{r}, \lambda \right )
\right ) 
L(\bm{r}) \label{subeq:Psi_C2},
\end{split}\\
\begin{split}
\Psi_D(\bm{\xi}, \lambda) &= \left (\frac{\lambda_0}{\lambda}\right )^2 \left (\widehat{\Psi}_A\left (\frac{\lambda_0}{\lambda}\bm{\xi}, \lambda \right ) \left (1 - e^{i\varphi_1}M(\bm{\xi}) - (e^{i\varphi_2} - e^{i\varphi_1})\,M_1(\bm{\xi}) - (1 - e^{i\varphi_2})\,M_2(\bm{\xi}) \right ) \right )\\
&*\widehat{L}\left (\frac{\lambda_0}{\lambda}\bm{\xi}, \lambda \right ). \label{subeq:Psi_D2}
\end{split}
\end{alignat}
\end{subequations}

In the absence of FPM, the electric field in the final image is given by
\begin{equation}
\Psi_D^0(\bm{\xi}, \lambda)= \left (\frac{\lambda_0}{\lambda}\right )^2 \widehat{\Psi}_A\left (\frac{\lambda_0}{\lambda}\bm{\xi}, \lambda \right )*\widehat{L}\left (\frac{\lambda_0}{\lambda}\bm{\xi}, \lambda \right ). 
\label{subeq:Psi_0}
\end{equation}

In the case of a flat stellar spectrum, the coronagraphic image intensity over $\Delta\lambda$ is given by
\begin{equation}
\mathcal{I}_{\Delta\lambda}[\Phi(\bm{r})](\bm{\xi})=\frac{1}{\Delta\lambda}\int_{\Lambda} \left |\Psi_D(\bm{\xi}, \lambda)\right |^2\,d\lambda\,,
\label{eq:I_Dbroad}
\end{equation}
where $\Lambda$ defines a set of wavelengths $\lambda$ such that $|\lambda-\lambda_0|< \Delta\lambda/2$.

Several families of solutions exist for the apodization to reduce the starlight diffracted by the aperture. Among them, we recently found broadband solutions that combines APLC with shaped pupil apodizations to achieve $10^{10}$ contrast with arbitrary pupils\cite{N'Diaye2015a,N'Diaye2016}. In this communication, we will focus on this kind of solutions and derive novel designs to push further the limits of the current designs in terms of IWA and spectral bands. 

\subsection{Throughput definitions}
Ruane et al. \cite{Ruane2017} presented different metrics to estimate the coronagraph performance. In this paper, we use two definitions for the throughput provided by the coronagraph as a first iteration:
\begin{itemize}
    \item transmitted energy (TE) throughput: also called coronagraph throughput, this term corresponds to the fraction of the residual energy within the Lyot stop in the absence of coronagraph mask with respect to the energy in the entrance pupil.
    \item encircled energy (EE) throughput: also defined as the core throughput\cite{Krist2016}, this term represents the fraction of the encircled energy for an off-axis companion within a radius of 0.7$\lambda_0/D$ at its location in the presence of the coronagraph without FPM (Apodizer and Lyot stop) with respect to the same amount in the absence of coronagraph (no Apodizer nor FPM or Lyot stop). As we here address the APLC case, the given value corresponds to the peak value obtained at all the separations for the off-axis companion beyond about the FPM projected radius ($\sim m/2 + 0.5\lambda_0/D$) in the final image plane, 
\end{itemize}

\section{Methodology for the apodization}
\label{sec:methodology}
\subsection{Numerical optimization for the apodizer}
We derive apodizer solutions by relying on shaped-pupil type optimizations\cite{Vanderbei2003a,Kasdin2003,Carlotti2011,Vanderbei2012}. A numerical optimization problem is defined with a Lyot-style coronagraphic device that includes the geometries of the FPM and Lyot Stop and we set constraints on the coronagraphic intensity in a region of interest with a given contrast $C$ over a spectral bandwidth $\Delta\lambda$. Among the possible apodizer solutions for the APLC, we search for the apodization with the best throughput in an attempt to maximize the number of photons from a planetary signal. The optimal solution is found by solving the following optimization problem:
\begin{subequations}
      \renewcommand{\theequation}{\theparentequation\alph{equation}}
\begin{alignat}{2}
& \max_{C} \left [\int_{P_0} \Phi(\bm{r})d\bm{r} \right] \quad \label{subeq:maxintP}\\
& \text{under the constraints:} \nonumber\\
& \left |\Psi_D(\bm{\xi}, \lambda)\right | < 10^{-C} \left |\Psi^0_D(\bm{0}, \lambda)\right | \quad \text{for} \quad
\begin{dcases}
\rho_0 < \xi < \rho_1\\
|\lambda-\lambda_0| < \Delta\lambda/2 \label{subeq:C} 
\end{dcases}\\
& 0 \leq \Phi(\bm{r}) \leq 1 \label{subeq:maxP},\\
\end{alignat}
\end{subequations}
Eqs. (\ref{subeq:maxintP}), (\ref{subeq:C}), and (\ref{subeq:maxP}) respectively represent the cost function, here the integrated apodizer transmission, the constraints on the coronagraphic electric field for a dark region ranging between $\rho_0$ and $\rho_1$ over the spectral band $\Delta\lambda$, and finally the constraints of positivity and normalization on the apodizer transmission.

In our previous communications, we used the AMPL algebraic modelling language\cite{Fourer1990} to write and convert the problem in a canonical form with matrices. However, we have recently tested python and found ways to compute the problem more efficiently by avoiding the introduction with lines with only null terms in the matrices. As this problem shows linear properties, solutions are found by using a linear programming (LP) solver. In our approach, we rely on the use of the Gurobi solver \cite{gurobi} which efficiently handles this type of problem. 

In our previous papers, we presented $10^{10}$ contrast solutions for circular axi-symmetric apertures\cite{N'Diaye2015a}. We then showed the existence of solutions for a LUVOIR-like segmented aperture that enable to produce a $10^{10}$ contrast PSF dark zone over a 10\% bandwidth, using a 4$\lambda_0/D$ radius FPM. We also show that some of these solutions are robust to the stellar angular size by producing a PSF core that proves smaller than the FPM\cite{N'Diaye2016}. 

Such designs offer somewhat moderate IWAs, bandwidth, and throughput, thus possibly limiting the number of observable Earth-like planets with future large missions for statistical studies. In this communication, we momentarily come back to the circular axi-symmetric aperture studies by performing 1D optimization problems as a first step to derive a new class of APLC designs.

\subsection{Combined optimization of the apodizer, FPM, and Lyot stop}
\subsubsection{Nelder-Mead optimization with embedded LP method}
By solving our optimization problem, we find APLC solutions with the optimal apodizer for a set of FPM and Lyot stop geometries to produce a given PSF dark zone. However, this linear problem does not encompass the simultaneous optimization of the FPM and Lyot Stop. We previously performed a full exploration of the parameter space to find coronagraph designs with the highest throughput\cite{N'Diaye2015a}. To search for all the optimal APLC parameters at once, we explore an alternative method based on a Nelder-Mead approach to find the best mask and diaphragm that maximize the EE throughput while using an embedded LP method to determine the optimal apodization for the FPM and Lyot stop parameters at each iteration of the process. 

\subsubsection{Parameter space exploration using MCMC approach}
We formerly explored the parameter space of FPM and Lyot stop dimensions to determine the optimal set of these parameters with the apodizer\cite{N'Diaye2015a}. This approach presents the advantage of mapping the space of solutions for further trade-offs on the coronagraph design between IWA, bandwidth, and throughput. In this work, we perform a MCMC approach using the throughput as a criteria to perform a parameter space exploration and analyze the space of solutions. The MCMC approach can be seen as an extension of the Nelder-Mead optimization approach if the same optimization criteria is used with both approaches. More details on this approach can be found in Fogarty et al. this conference\cite{fogarty2018}. 

\section{Application to large telescope apertures}
\label{sec:application}
\subsection{Assumptions}
In this study, we assume a circular axi-symmetric aperture and therefore, we design our coronagraphs using one-dimension formalism. Within this framework, the electric field in the different planes can be expressed by simply replacing the 2D Fourier transform with the 1D Hankel transform in the equations that are given in section \ref{sec:principle}. More details on the analytical form can be found in the literature \cite{Pueyo2013,N'Diaye2015a}.

We remind the reader that $m$, $ID$ and $OD$ denote the diameter of the FPM, inner and outer Lyot stop while $R_{FPM}$, $R_{ILS}$, and $R_{OLS}$ here represent their respective radii. Considering a 14\% central obstruction for the aperture, we aim to find solutions that produce a $10^{10}$ contrast dark zone ranging between 2.5 and 10$\lambda/D$ from the star in the coronagraphic image. Unless otherwise stated, we adopt a 20\% spectral bandwidth, covering a wide spectral range.  

In our 2015 paper\cite{N'Diaye2015a}, we looked for continuous solutions by setting constraints on the derivative of the apodizer transmission. In these preliminary studies, we only focus on binary solutions and therefore, no derivative constraints is included here. 

\subsection{Small IWA solutions with opaque FPMs for large spectral bands}
\begin{figure}[!ht]
\centering
\includegraphics[width=0.45\textwidth]{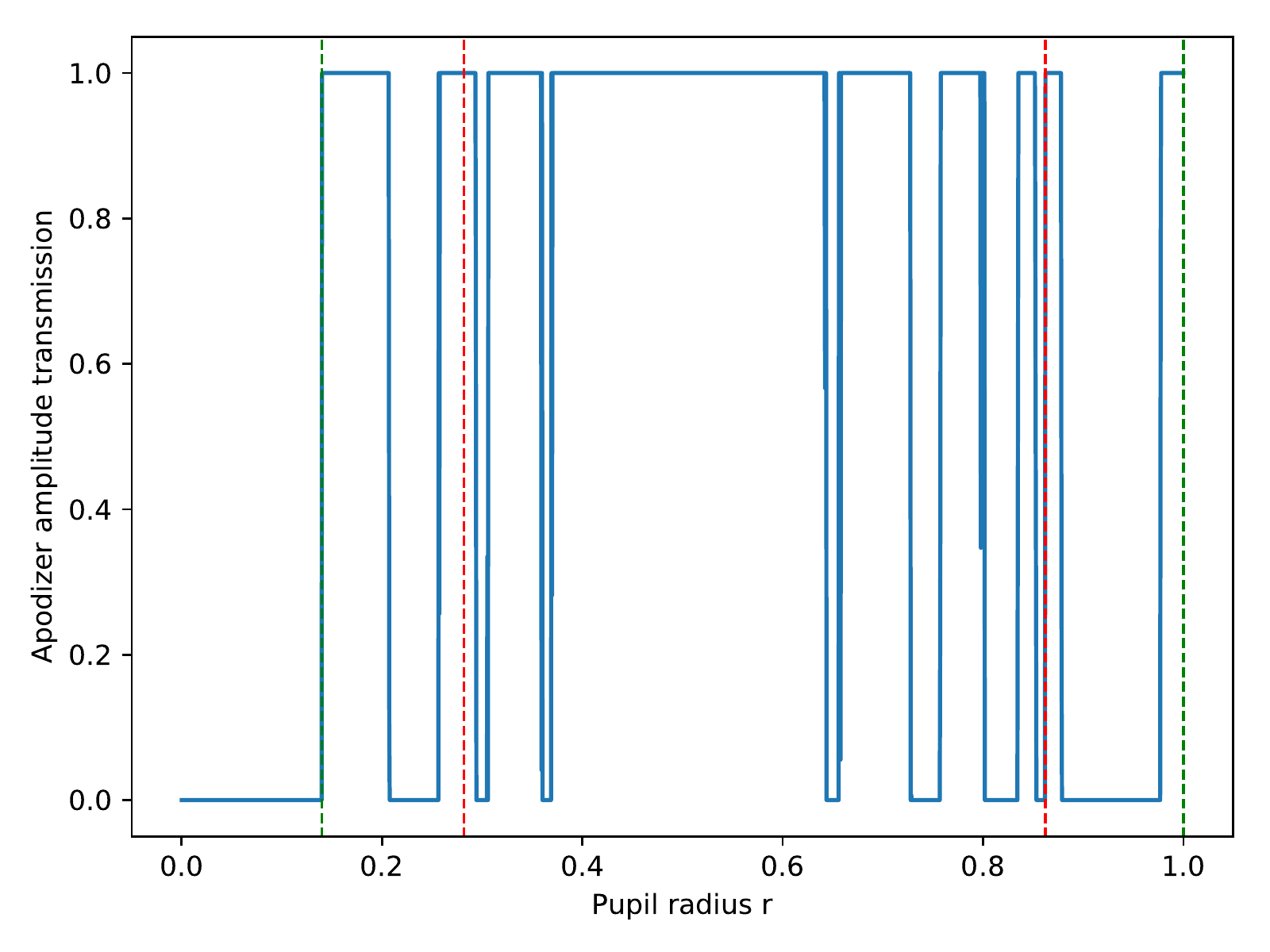}
\includegraphics[width=0.45\textwidth]{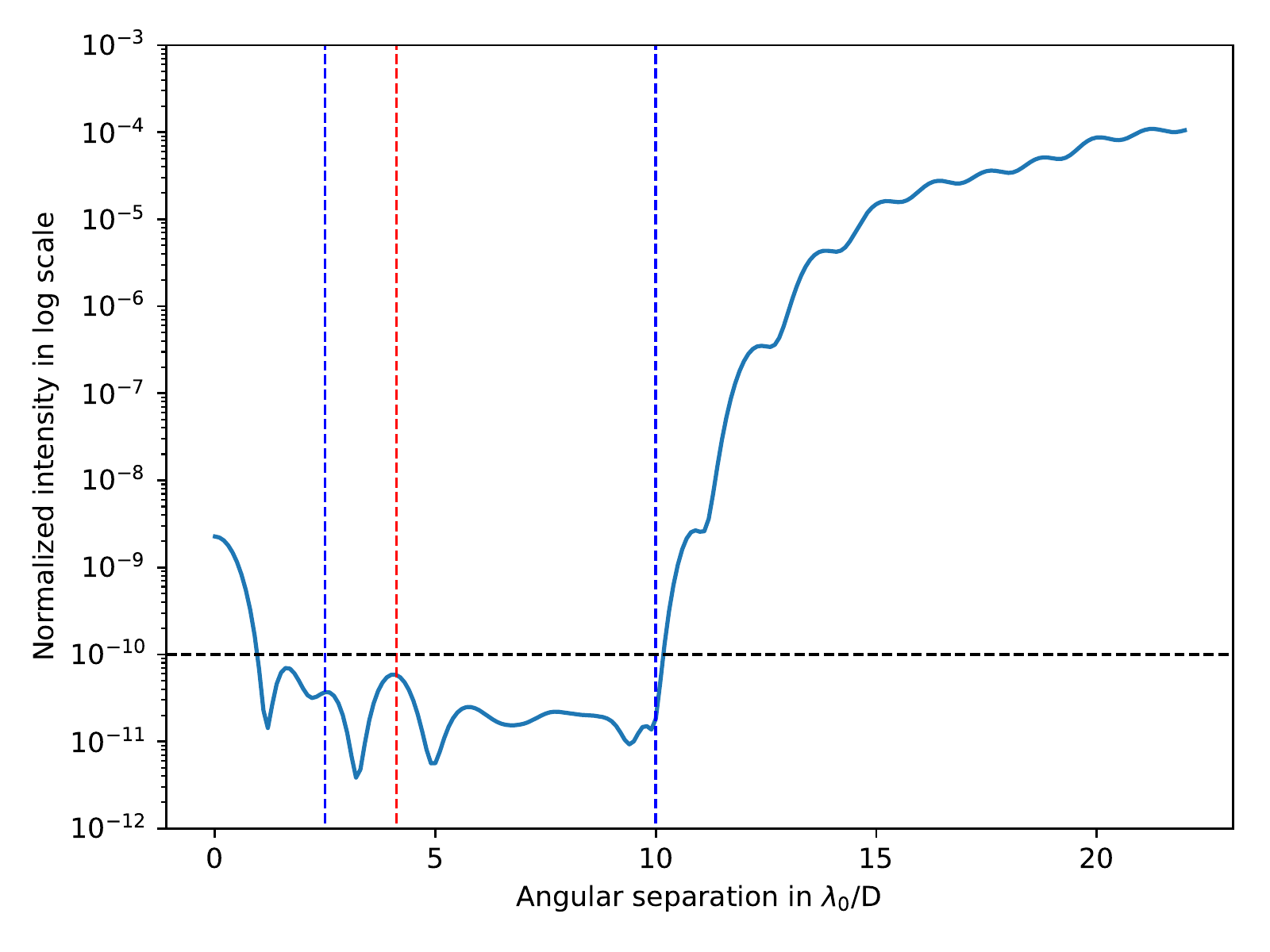}
\includegraphics[width=0.45\textwidth]{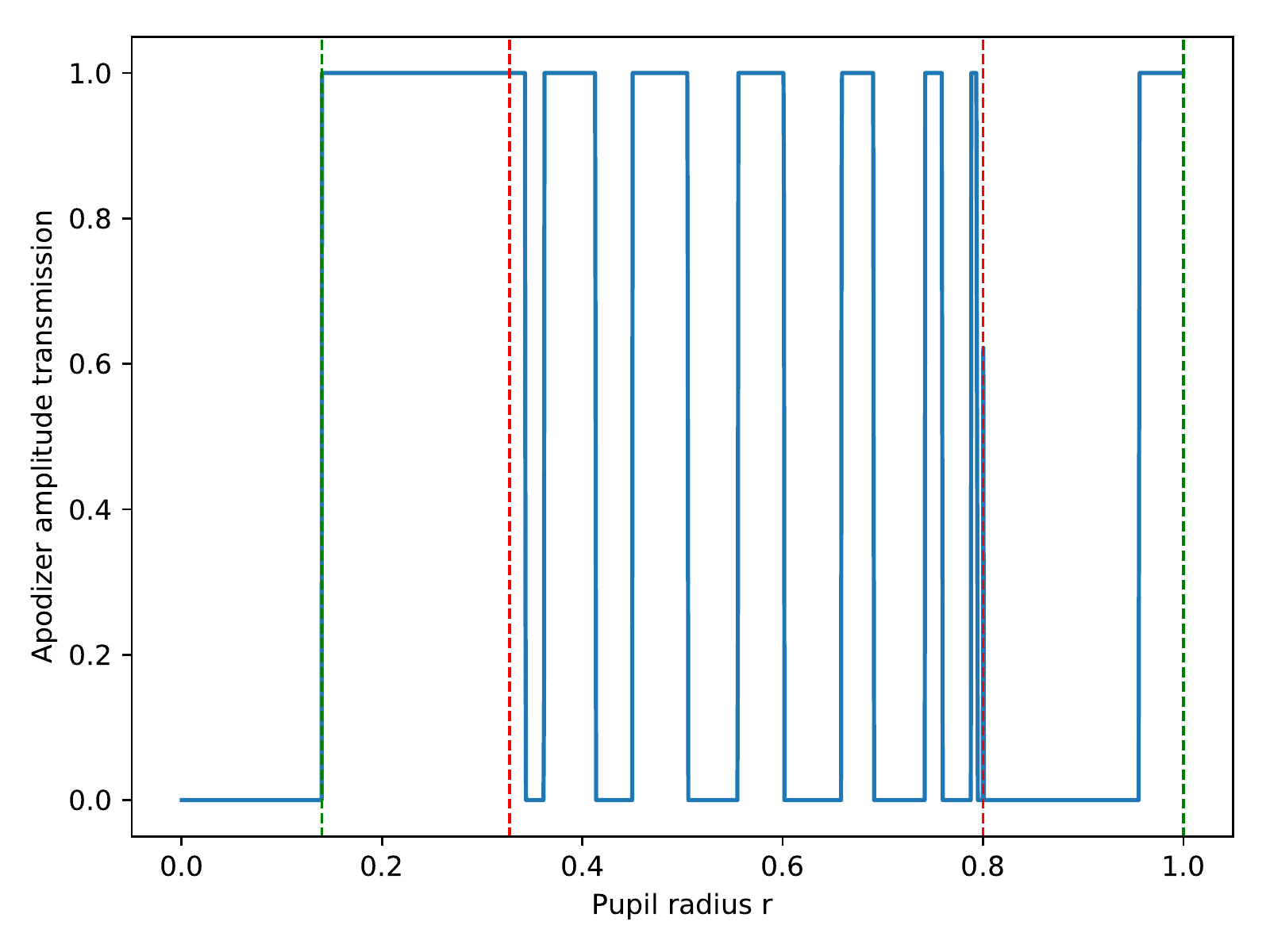}
\includegraphics[width=0.45\textwidth]{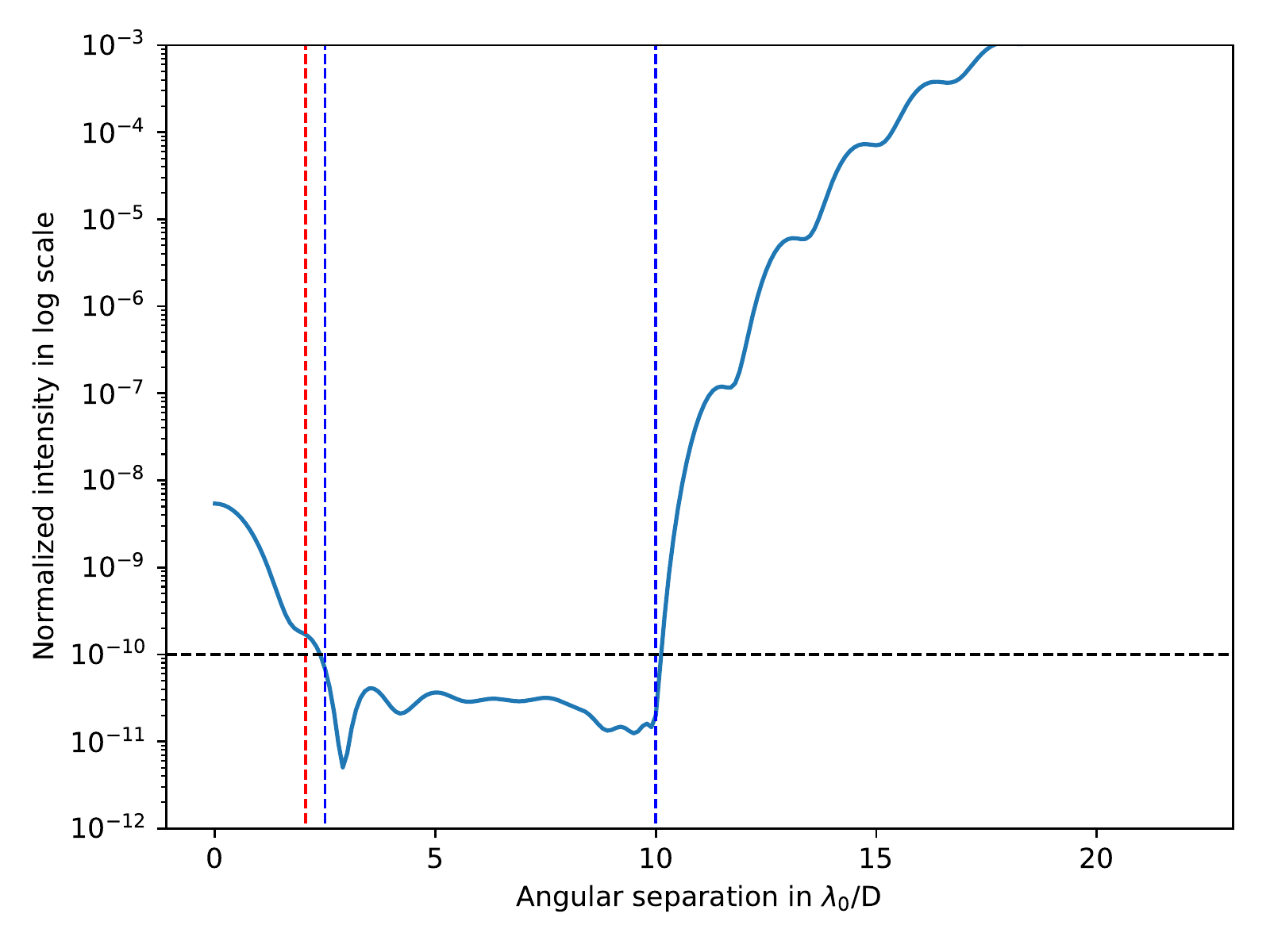}
\caption{APLC solutions for a 14\% centrally obstructed circular aperture to produce a star image with a $10^{-10}$ residual starlight intensity ranging between 2.5 and 10$\lambda_0/D$ from an observed star over 20\%bandwidth. These designs are obtained for a opaque FPM with 4.12 and 2.06$\lambda_0/D$ radius (top and bottom). \textbf{Left}: Radial amplitude profile of the apodization transmission. The green and red vertical lines delimit the inner and outer boundaries of the entrance pupil and Lyot stop respectively. \textbf{Right}: Radial intensity profile of the coronagraphic image for the APLC solutions. The blue vertical lines delimit the star image high-contrast region and the red line denote the projected FPM in the final image plane.}
\label{fig:APLC_plots}
\end{figure}

Using the Nelder-Mead approach, we first explore the APLC solutions by considering the following ranges: $2 < m/2 < 4.5\lambda/D$, $0.14  < ID < 0,42$, and $0.8 < OD < 1.0$. Figure \ref{fig:APLC_plots} top plot shows our optimal solution and its apodization with $m/2 = 4.117\lambda_0/D$, $ID=0.282$, and $OD=0.862$, see Table \ref{tab:APLC}. The Lyot stop has a central obstruction diameter that is about twice larger than the aperture, showing good agreement with the previously found dimensions to obtain achromatic designs \cite{Soummer2011, N'Diaye2015a}. In addition, the Lyot Stop has an outer diameter slightly smaller than the aperture. As the light diffracted by the FPM is mostly located at the relayed pupil edge, a reduced Lyot stop allows for removing most of the star diffracted light. This solution displays a FPM radius of about 4\,$\lambda_0/D$, showing good consistency with the results of our previous works to achieve high transmission with our device. It is worth noting than the FPM radius is larger than the dark zone inner edge. Such a property tends to make our design more robust to stellar angular size or low-order aberrations which can originate from telescope jitter or thermal/mechanical focus drifts\cite{N'Diaye2015a}.

While promising, this first design presents a large effective IWA, approximately corresponding to the 4\,$\lambda_0/D$ FPM radius here. These solutions limit the discovery space of planets at the closest separations from an observed bright star. To overcome this issue, we look for solutions with smaller FPM by limiting the search range of the FPM with our Nelder-Mead approach: $2 < m/2 < 3.0\lambda/D$, and keep the same $ID$ and $OD$ ranges. Figure \ref{fig:APLC_plots} bottom plot shows our solution with $m/2=2.058\lambda_0/D$, $ID=0.327$, and $OD=0.800$, see Table \ref{tab:APLC}. This solution uses a smaller Lyot stop outer edge, mostly likely resulting from the presence of a smaller FPM which widely diffracts more light at the relayed pupil edge to compensate for. Although at the expense of a lower throughput (both TE and EE), these APLC solutions are interesting since they enable to observe companions around nearby stars over a 20\% bandwidth with an effective IWA smaller than 3.0$\lambda_0/D$. Such a solution was not detected in previous 1D studies. This is most likely due to the presence of additional and strong constraints on the derivative of the apodizer transmission.

\begin{table}[!ht]
    \centering
    \caption{Parameters for the optimal APLC solutions using opaque FPM to produce a star image with a $10^{10}$ contrast dark zone ranging between 2.5 and 10$\lambda_0/D$ from the star over a 20\% bandwidth.}
    \begin{tabular}{c c c c c c}
    \hline\hline
         Parameters/results & $m$/2 in $\lambda_0/D$ & ID & OD & TE in \% & EE in \%\\\hline
         Fig. \ref{fig:APLC_plots} top    plot & 4.117 & 0.282 & 0.862 & 81.2 & 30.4\\
         Fig. \ref{fig:APLC_plots} bottom plot & 2.058 & 0.327 & 0.800 & 47.1 &  6.5\\\hline
    \end{tabular}
    \label{tab:APLC}
\end{table}

\begin{figure}[!ht]
\centering
\includegraphics[width=0.45\textwidth]{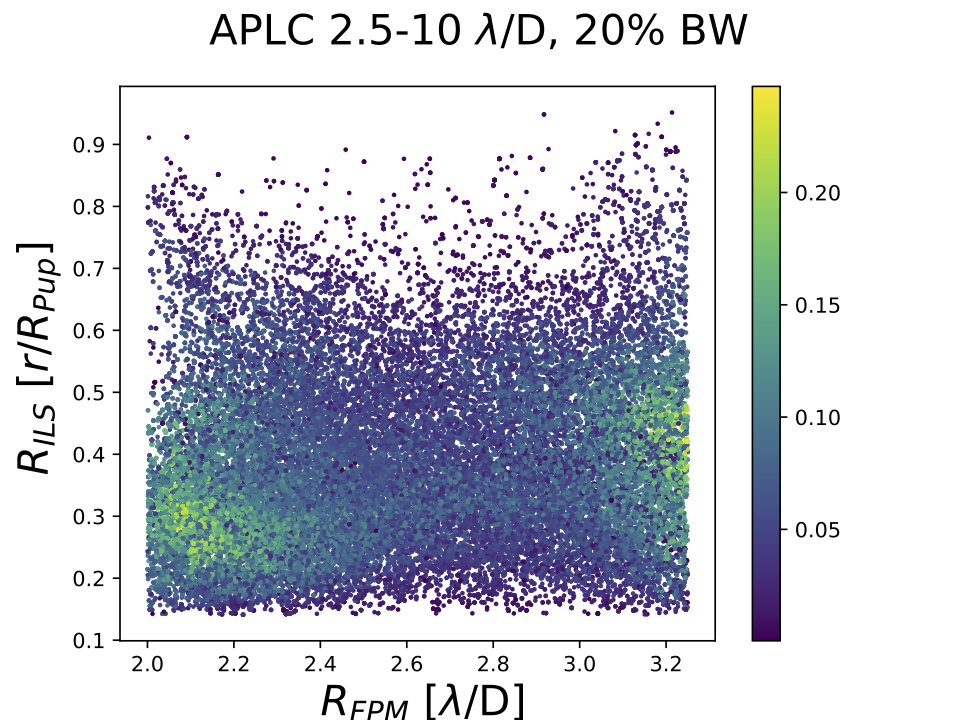}
\caption{A 2-D cut of the 3-D parameter space (FPM radius $R_{FPM}$, inner Lyot Stop radius $R_{ILS}$, Outer Lyot Stop radius $R_{OLS}$) sampled via MCMC for a 14\% centrally obstructed circular aperture. In this plot, we show the T.E. throughput as a function of FPM and inner Lyot Stop radii for $R_{OLS}=0.9$. We used 6 walkers which sampled the parameters space at $10^4$ points per walker, and optimized pupil apodizers to produce a star image with 2.5-10 $\lambda_0/D$ dark holes with $10^{-10}$ starlight residual intensity over 20\% bandwidth for each configuration of ($R_{FPM}$, $R_{ILS}$, $R_{OLS}$) sampled.}
\label{fig:APLC_contour_plots}
\end{figure}

To get a better sense on this existence of this solution, we decide to map the parameter space of solutions by using the MCMC approach, see Figure \ref{fig:APLC_contour_plots}. Each point in this plot represents a solution that is color-coded in EE throughput and the points are distributed along the FPM radius and inner Lyot stop radius for a given Lyot stop outer edge of $0.9$. In this plot, we clearly observe two zones of relatively high transmitted solutions. The right-hand side zone corresponds to the well-known APLC solutions with FPM larger than $3\lambda_0/D$. However, the left-hand side of the plots shows solutions with small FPM, corresponding to the solution showed in Figure \ref{fig:APLC_plots} bottom plot. Both sets of solution are separated by a discontinuum with a FPM radius around 2.7$\lambda/D$, which corresponds to the location of the second bright diffraction ring for our considered aperture. The throughput of our solutions seems to be somewhat related to the radius of the FPM with respect to the location of the bright and dark diffraction rings. Further investigations will be performed to confirm this existence of this property. A more in-depth study of this parameter space of solutions is presented by Fogarty at al. in these proceedings\cite{fogarty2018}. All in all, this presence of this parameter space for APLC solutions with small IWA opens up prospects for close-in observations with such a device.

\section{Conclusion}
\label{sec:conclusion}
We have revisited the APLC concept by exploring novel optimization schemes to identify new solutions with smaller IWA ($< 3\lambda_0/D$). For the circular axi-symmetric apertures, we have found designs to form the image of an observed star with a $10^{10}$ contrast at separations ranging between 2.5 and 10\,$\lambda_0/D$ over 20\% spectral bandwidth. These solutions open up a new window for the study of close-in planets with the APLC, in particular with the observation of Earth-like planets with future large missions. 

Our preliminary designs constitute a proof of concept on the existence of additional solutions for the APLC to simultaneously achieve small IWA, deep contrast, high throughput, and wide spectral band. In this work, we have adopted the EE throughput as a criteria to find our solutions. However, further studies should include exoplanet yield\cite{Stark2014,Stark2015} and design reference mission aspects \cite{Brown2005,Savransky2016} as a criteria to maximize the number of observable targets with high-contrast instruments on future large space observatories and enable statistical studies on rocky planets.

Relying on the existing DZPM concepts \cite{Soummer2003b,N'Diaye2012a,Delorme2016}, we have revisited the APLC by replacing the traditional opaque mask with an alternative hybrid mask that combines an opaque dot with two annular phase shifting zones to reduce the IWA further. We have set a generalized formalism of the APLC with hybrid FPMs to explore novel concepts and reduce the APLC IWA further. In the near future, we will investigate new APLC designs with an hybrid mask with applications for future missions with exo-Earth imaging capabilities but in the current ground-based exoplanet imagers, such as GPI, SPHERE, or SCExAO \cite{Beuzit2008,Macintosh2014,Jovanovic2015}, with a possible upgrade offering minimal system modifications in the current ground-based exoplanet imagers using opaque masks.

Our current analysis is limited to coronagraphic designs for a circular aperture with central obstruction. However, these static solutions can be combined with active control of aperture discontinuities (ACAD) methods\cite{Pueyo2013,Mazoyer2018a,Mazoyer2018b} to address the diffraction features due to spiders or pupil segmentation in a telescope primary mirror, and enable the formation of PSF dark zones with any arbitrary telescope apertures. In this venue, much work is in progress and the first results can be found in Fogarty et al. this conference. In a further work, we will alternatively investigate static solutions that can handle all the features of any arbitrary telescope aperture. 

All these aspects will prove helpful for further studies on complete systemic approaches in the context of future large segmented aperture telescopes such as LUVOIR or HabEx mission concepts to observe exo-Earths. Ground-based observatories can also benefit from these starlight suppression systems for further upgrades of the current exoplanet imagers to observe fainter or closer-in planetary companions or in preparation for the future ELTs with the detection of rocky planets around M dwarfs.   

\acknowledgments 
M.N. acknowledges support from Action Sp\'ecifique Haute R\'esolution Angulaire (ASHRA) and his home institution Laboratoire Lagrange. This work is partially supported in part by the National Aeronautics and Space Administration under Grants NNX12AG05G and NNX14AD33G issued through the Astrophysics Research and Analysis (APRA) program (PI: R. Soummer) and by Jet Propulsion Laboratory subcontract No.1539872 (Segmented-Aperture Coronagraph Design and Analysis; PI: R. Soummer).
% References
\bibliography{2018_mndiaye_biblio_v1}
\bibliographystyle{spiebib_5authors}

\end{document}